\documentclass[11pt]{jpp}

\newcommand{\be}{\begin{displaymath}}
\newcommand{\bn}{\begin{equation}}
\newcommand{\bea}{\begin{eqnarray*}}
\newcommand{\eea}{\end{eqnarray*}}
\newcommand{\en}{\end{equation}}
\newcommand{\ee}{\end{displaymath}}

\renewcommand{\p}{\partial}
\newcommand{\lang}{\left\langle}
\newcommand{\rang}{\right\rangle}

\newcommand{\simgt}{\:{\raisebox{-1.5mm}{$\stackrel
{\textstyle{>}}{\sim}$}}\:}
\newcommand{\simlt}{\:{\raisebox{-1.5mm}{$\stackrel
{\textstyle{<}}{\sim}$}}\:}

\shorttitle{Electron-positron plasmas}
\shortauthor{P. Helander and J. W. Connor}

\title{Gyrokinetic stability theory of electron-positron plasmas}

\author{Per Helander\aff{1}
  \corresp{\email{per.helander@ipp.mpg.de}}
 \and J. W. Connor\aff{2}}

\affiliation{\aff{1}Max-Planck-Institut f\"ur Plasmaphysik, 17491 Greifswald, Germany
\aff{2}Culham Centre for Fusion Energy, Abingdon, OX14 3DB, United Kingdom}

\begin{document}

\maketitle

\begin{abstract}
The linear gyrokinetic stability properties of magnetically confined electron-positron plasmas are investigated in the parameter regime most likely to be relevant for the first laboratory experiments involving such plasmas, where the density is small enough that collisions can be ignored and the Debye length substantially exceeds the gyroradius. 
Although the plasma beta is very small, electromagnetic effects are retained, but magnetic compressibility can be neglected. The work of a previous publication (\cite{Helander-2014})
is thus extended to include electromagnetic instabilities, which are of importance in closed-field-line configurations, where such instabilities can occur at arbitrarily low pressure. 
It is found that gyrokinetic instabilities are completely absent if the magnetic field is homogeneous: any instability must involve magnetic curvature or shear. 
Furthermore, in dipole magnetic fields, the stability threshold for interchange modes with wavelengths exceeding the Debye radius coincides with that in ideal MHD. 
Above this threshold, the quasilinear particle flux is directed inward if the temperature gradient is sufficiently large, leading to spontaneous peaking of the density profile. 
\end{abstract}

%\begin{center}
%\Large
%{\bf Gyrokinetic stability theory of electron-positron plasmas}
%~\\
%\normalsize
%\vspace{1cm}
%P. Helander$^1$ and J.~W.~Connor$^{2}$
%\\[1cm]
%{\it $^1$Max-Planck-Institut f\"ur Plasmaphysik, 17491 Greifswald, Germany \\
%$^2$ Culham Centre for Fusion Energy, Abingdon, OX14 3DB, United Kingdom}
%\vspace{0.5cm}
%\end{center}
%\noindent
%\noindent
%\newpage 

\section{Introduction}

An experiment aiming at creating the first man-made electron-positron plasma has recently been reported in 
%{\em Nature Communications} 
\cite{Sarri-2015}. The positrons were created through the collision of a laser-produced electron beam with a lead target, and the experiment represents a major step in the quest to study the physics of electron-positron plasmas. These are, of course, the simplest plasmas imaginable and play an important role in astrophysics and cosmology. 

Laser-produced plasmas are necessarily highly nonstationary, but other experiments are underway that attempt to create steady-state electron-positron plasmas confined by magnetic fields. The idea is to use a nuclear reactor to create a large number of positrons, accumulate these in a Penning trap, and inject them into a confining magnetic field with the geometry of either a simple stellarator or a magnetic dipole (\cite{Pedersen-2003,Pedersen-2012}). The first attempts to inject positrons into a dipole field have already proved successful (\cite{Saitoh-2015}). The aim is to produce a quasineutral plasma with a positron number density in the range $10^{12} \, {\rm m}^{-3} < n < 10^{13} \, {\rm m}^{-3}$ and a temperature $T$ between 1 and 10\,eV. The Debye length $\lambda_D = (\epsilon_0 T / 2 ne^2)^{1/2}$ then becomes as short as a few mm, significantly smaller than the diameter of the plasma. On the other hand, the Debye radius exceeds the gyroradius by two to three orders of magnitude if the magnetic field is of the order of $B=1$\,T.
 
If the density and temperature of the electrons and positrons are equal, there is perfect symmetry between the positive and negative species, and a number of phenomena familiar from conventional plasmas disappear (\cite{Tsytovich-1978}). There are, for instance, no drift waves or Faraday rotation, and sound waves are strongly Landau damped, because their phase velocity is equal to the thermal speed of both species. As a result, those microinstabilities that normally arise because of the destabilization of sound waves or drift waves due to cross-field gradients are absent. There is, for instance, no analog to the ``slab'' ion-temperature-gradient mode. In fact, any instability must involve magnetic curvature. These results follow from a simple stability analysis based on the gyrokinetic equation in the electrostatic limit that was carried out in %Ref.~
\cite{Helander-2014}, where, in addition, the stability against electromagnetic modes was estimated on the basis of MHD interchange instability. A proper treatment should, of course, address both types of modes within the gyrokinetic framework. This is the subject of the present paper. 

Since the collision frequency for the parameters given above is smaller than the expected mode frequencies, we neglect Coulomb collisions,  in contrast to the work by %Simakov, Hastie and Catto 
\cite{Simakov-2002}, who pioneered the study of gyrokinetic instabilities in dipole fields, but considered conventional ion-electron plasmas. We note that direct electron-positron annihilation can be neglected too (\cite{Helander-2003}). The cross section for annihilation of positrons in collisions with stationary electrons is (\cite{Heitler-1953})
	\bn 
	  \sigma_a = \frac{\pi r_e^2}{1 + \gamma} \left[ \frac{\gamma^2 + 4 \gamma + 1}{\gamma^2-1} 
	  \ln \left(\gamma + \sqrt{\gamma^2 -1} \right) - \frac{\gamma + 3}{\sqrt{\gamma^2 - 1}} \right]
	\label{annihilation}
	\en
where $r_e = e^2 / 4 \pi \epsilon_0 m_e c^2$ is the classical electron radius and $\gamma$ the Lorentz factor. If the energy of the positron is low enough, this cross section must be corrected to account for the focusing effect of Coulomb attraction between the electron and the positron, and then becomes 
(\cite{Baltenkov-1985})
	$$ \sigma_a \rightarrow \frac{\sigma_a \xi}{1 - e^{-\xi}}, 
	$$
where $\xi = 2 \pi \alpha c/v$ and $\alpha \simeq 1/137$ denotes the fine-structure constant. This correction is important if $\xi \simgt 1$, i.e., if $E = m_e v^2 / 2 \; \simlt \; 500$\,eV. The resulting annihilation frequency, which becomes
	$$ \nu_a = n v \sigma_a \simeq 2 \alpha n v \left( \frac{\pi c r_e}{v} \right)^2 
	$$
at low energies, is nevertheless much smaller than the Coulomb collision frequency $\nu_e$, by a factor
	$$ \frac{\nu_a}{\nu_e} \sim \left( \frac{v}{c} \right)^2 \alpha \ln \Lambda, 
	$$
where $\ln \Lambda$ denotes the Coulomb logarithm. 

In fact, the primary loss mechanism for positrons will not be direct annihilation with electrons, but is more likely to be caused by positronium formation in charge-exchange reactions with neutral atoms 
(\cite{Helander-2003}). Positronium is highly unstable and has a life time shorter than 1\,$\mu$s. 

Coulomb	collisions are important in the sense that they ensure that the equilibrium distribution function is Maxwellian, since the collision frequency exceeds the expected inverse confinement time by a large factor. They may also affect instabilities if the frequency $\omega$ of the latter is smaller than, or comparable to, the collision frequency, i.e., if $\omega \simlt \nu_e$, but this will not be the case for most instabilities of interest.

\section{Gyrokinetic system of equations}

An electron-positron plasma with the parameters quoted above has a very low value of 
$\beta = 4 \mu_0 nT/B^2$, but we shall nevertheless retain electromagnetic effects in the gyrokinetic equation,
	\bn 
	  i v_{\|}\nabla_{\|}g_a+(\omega -\omega_{da})g_a = \frac{e_a}{T_a} J_0 
	  \left(\frac{k_\perp v_\perp}{\Omega_a} \right)  
	  \left( \omega - \omega_{*a}^T \right)(\phi - v_\| A_\| ) f_{a0}, 
	\label{gk}
	\en
where $\Omega_a = e_a B / m_a$ denotes the gyrofrequency, ${\bf k}_\perp = k_\psi \nabla \Psi + k_\varphi \nabla \varphi$ the wave vector perpendicular to the magnetic field, ${\bf B} = \nabla \Psi \times \nabla \varphi$ the equilibrium magnetic field, $J_0$ a Bessel function, $\omega_{*a}^T = \omega_{*a} \left[ 1 + \eta_a \left(m_a v^2 / 2T - 3/2 \right) \right]$ with $\eta_a = d \ln T_a / d \ln n_a$, $\omega_{*a} = (T_a k_{\varphi}/e_a) d \ln n_a / d \Psi$ the diamagnetic frequency, and $\omega_{da} = {\bf k}_\perp \cdot {\bf v}_{da}$ the magnetic drift frequency. The independent variables are the velocity $v$ and the magnetic moment $\mu = v_\perp^2 / (2 B)$. The electrostatic potential is denoted by $\phi$, and the component along $\bf B$ of the magnetic potential $\bf A$, in the Coulomb gauge ($\nabla \cdot {\bf A} = 0$), is denoted by $A_\|$. As we shall see, the retention of $A_\|$ is necessary in order to calculate stability boundaries even for cases with $\beta \ll 1$. Magnetic compressibility can however be neglected, i.e., we may take $\delta B_\| = 0$, as long as $\beta$ is small. In Eq.~(\ref{gk}), the distribution function has been written as 
	$$ f_a({\bf r}, t) = f_{a0} ({\bf r},H) \left( 1 - \frac{e_a \phi({\bf r},t)}{T_a} \right) 
	  + g_a({\bf R},H,\mu,t), 
	$$
where ${\bf R} = {\bf r} + {\bf b} \times {\bf v} / \Omega_a$ denotes the guiding-centre position, $f_{a0}$ the Maxwellian, $H = m_a v^2 / 2 + e_a \phi({\bf r},t)/T_a$ the energy and ${\bf b} = {\bf B}/B$. The coordinate $\Psi$ has been chosen such that $f_{a0}$ is constant on surfaces of constant $\Psi$, and the coordinate $\varphi$ thus labels the different field lines on each such surface. In the case of a dipole field, which will be treated explicitly below, $\varphi$ denotes the azimuthal angle. 

The gyrokinetic equation~(\ref{gk}) holds both in magnetic geometries with closed field lines, and in geometries where the magnetic field lines do not close on themselves and there is magnetic shear. In the latter case, one must employ the ballooning transform, but in the following, we need not distinguish between the two cases. 

The fields $\phi$ and $A_\|$ are determined by Poisson's and Amp\`ere's laws, respectively, which in the present notation become
	$$ \left( \sum_a \frac{n_a e_a^2}{T_a} + \epsilon_0 k_\perp^2 \right) \phi = \sum_a e_a \int g_a J_0 \; 
	   d^3v, 
	$$
	\bn 
	  A_\| = \frac{\mu_0}{k_\perp^2} \sum_a e_a \int v_\| g_a J_0 \; d^3v. \label{delta A} 
	\en
When discussing symmetric electron-positron plasmas, we shall frequently suppress the species subscript, understanding it to refer to the positrons whenever there is no risk of misunderstanding. The field equations can thus be written as
	\bn 
	  \left( 1 + k_\perp^2 \lambda_D^2 \right) \phi = \frac{T}{2ne} \int \left( g_p - g_e \right) J_0 d^3v, 
	\label{phi}
	\en
	\bn 
	  A_\| = \frac{\mu_0 e}{k_\perp^2} \int v_\| ( g_p - g_e) J_0 d^3v, 
	\label{A}
	\en
where $\lambda_D^2 = \epsilon_0 T / (2ne^2)$ denotes the square of the Debye length. Our complete system of equations is thus furnished by Eqs.~(\ref{gk}), (\ref{phi}) and (\ref{A}).

\section{Straight magnetic field}

This system of equations is trivially solved in the case of a constant magnetic field, where we can take a Fourier transform in the coordinate $l$ along $\bf B$, and find
	$$ g_{p,e} = \frac{\pm \omega - \omega_\ast^T}{\omega - k_\| v_\|} \frac{e J_0}{T} \; 
	   \left( \phi - v_\| A_\| \right) f_0. 
  $$
A remarkable conclusion follows immediately from the fact that neither the density nor temperature gradient appears in the quantity 
	\bn 
	  g_p - g_e = \frac{2 \omega}{\omega - k_\| v_\|} \frac{e J_0}{T} \; \left( \phi - v_\| A_\| \right) f_0 
	\label{difference}
	\en
that enters in the field equations (\ref{phi}) and (\ref{A}). Hence it is clear that the linear dynamics is similar to that of a homogeneous plasma. In particular, we conclude that {\em no gyrokinetic instabilities are present in an electron-positron plasma embedded in a constant magnetic field}. That this is true in the electrostatic case ($A_\| = 0$) was shown in %Ref.~
\cite{Helander-2014}, and now we see that it is more generally true. As long as the orderings underlying gyrokinetics are satisfied, no instability is possible, no matter how large the gradients are. The situation is similar to ideal magnetohydrodynamics, which is always stable in a constant magnetic field. Any instability must involve magnetic curvature or shear. 

The physical reason for this result becomes apparent if one analyses the waves of the system. Substituting the solution (\ref{difference}) into the field equations (\ref{phi}, \ref{A}) and evaluating the integrals
gives 
	$$ \left[ 1 + k_\perp^2 \lambda_D^2 + x Z(x) \Gamma_0(b) \right] \phi - \Gamma_0(b) x \left[1 
	   + x Z(x)   \right] v_T A_\| = 0, 
	$$
	$$ \Gamma_0(b) x \left[1 + x Z(x) \right] \phi - \left[ x^2 (1 + x Z(x)) \Gamma_0(b) - a \right] v_T A_\| 
	   = 0, 
	$$
where $x = \omega / (k_\| v_T)$, $v_T = \sqrt{2T/m}$ denotes the thermal velocity, $b = k_\perp^2 T / m \Omega^2 = k_\perp^2 \rho^2$, $\Omega = eB/m$, $\Gamma_0(b) = I_0(b) e^{-b}$, and 
	$$ a = \frac{k_\perp^2 m}{4 \mu_0 n e^2} = \left( \frac{k_\perp \lambda_D c}{v_T} \right)^2. 
	$$
The dispersion relation is obtained by equating the determinant of this system of equations to zero, giving 
	\bn 
	  \left(1 - \Gamma_0 + k_\perp^2 \lambda_D^2\right)  \left[\Gamma_0 x^2 (1+xZ) - a \right] 
		- a \Gamma_0 (1+xZ) = 0, 
	\label{disprel}
	\en
which is a special case of the general gyrokinetic dispersion relation in a homogeneous plasma given by %Howes et al 
\cite{Howes-2006}. Note that, since the equilibrium gradients have dropped out, there are no drift waves and therefore no drift instabilities, regardless of the size of the density and temperature gradients. 

For the plasmas we are primarily interested in, the Debye length exceeds the gyroradius, 
$\lambda_D \gg \rho$, and the thermal speed is much smaller than the speed of light, so that $a/(k_\perp \lambda_D)^2 = (c/v_T)^2 \gg 1$. The dispersion relation (\ref{disprel}) then reduces to that of ordinary electrostatic sound waves, 
	$$ 1+xZ(x) + k_\perp^2 \lambda_D^2 = 0, 
	$$
which are heavily Landau-damped if $k_\perp \lambda_D$ is of order unity or larger. Electromagnetic waves are thus absent in this limit, and the reason for the absence is easily understood by noting that the ratio 
$\lambda_D/\rho$, which we have taken to be much larger than unity, can also be written as 
	$$ \frac{\lambda_D}{\rho} =  \frac{v_A}{c}, 
	$$
where $v_A = B / \sqrt{2 m n \mu_0}$ denotes the Alfv\'en speed. Thus, in the present gyrokinetic ordering, which only considers waves that are much slower than light, Alfv\'en waves cannot exist because $v_A > c$. These are instead obtained from Eq.~(\ref{disprel}) in the limit $k_\perp \rho \sim k_\perp \lambda_D \ll 1$, where the first term containing $a$ is negligible compared with the second one, so that
	$$ x^2 = \frac{a}{k_\perp^2 \lambda_D^2 + b}, 
	$$
i.e.,
	$$ \frac{\omega}{k_\|} = \frac{v_A}{\sqrt{1 + v_A^2 / c^2}}. 
	$$
	
To summarise this section, we conclude that there are no linear gyrokinetic instabilities in a pair plasma embedded in a constant magnetic field, regardless of the size of the density and temperature gradients. In particular, there are no drift instabilities because of the absence of simple drift waves, and there are no unstable versions of sound waves (which are Landau damped) or Alfv\'en waves, which are absent if $\rho < \lambda_D$. 

\section{Curved magnetic field}

\subsection{Governing equations}

Gyrokinetic instabilities are thus only possible in curved (or sheared) magnetic fields, which we now consider. Equation~(\ref{gk}) is easily solved in the limit where the first term is either very small or very large. In the first case, we take $\omega \gg v_T/L$, where $L$ denotes the macroscopic length of the plasma, and obtain
	\bn 
	  g_{p,e} = \frac{\pm \omega - \omega_\ast^T}{\omega \mp \omega_d} \frac{e J_0}{T} \; 
		\left( \phi - v_\| A_\| \right) f_0. 
	\label{low-omega}
	\en
The equation (\ref{phi}) then becomes identical to that in the electrostatic limit (\cite{Helander-2014}), 
	$$ 1 - \Gamma_0(b) + k_\perp^2 \lambda_D^2 
	   = \frac{1}{n} \int \frac{\omega_d (\omega_d - \omega_\ast^T)}{\omega^2 - \omega_d^2} J_0^2 f_0 d^3v, 
	$$
but the frequency predicted by this dispersion relation will necessarily contradict the assumption $\omega \gg v_T/L$, because this assumption implies that the right-hand side of the equation is at most of order $k_\perp^2 \rho^2$ whereas the left-hand side is of order $k_\perp^2 \lambda_D^2$, which is much larger if $\rho \ll \lambda_D$. 

As in %Ref.~
\cite{Helander-2014}, we are thus led to consider the opposite limit, $\omega \ll v_T/L$. Ignoring reconnecting instabilities and following %Tang et al.~
\cite{Tang-1980}, we write
	$$ A_\| = \frac{\nabla_\| \psi}{i \omega}. 
	$$
and expand the distribution function in the small parameter $\omega L / v_T \ll 1$. In lowest order we obtain
	$$ g_a = \left( 1 - \frac{\omega_{\ast a}^T}{\omega} \right) \frac{e_a \psi}{T_a} f_{a0} + h_a, 
	$$
where the integration constant satisfies $\nabla_\| h_a = 0$ and is determined by the integrability condition of the next-order equation, 
	$$ \overline {\left( \omega - \omega_{da} \right) \left[\left( 1 - \frac{\omega_{\ast a}^T}{\omega} \right) 
	   \frac{e_a \psi}{T_a} f_{a0} + h_a \right] }
	   = \left( \omega - \omega_{\ast a}^T \right) \frac{e_a \overline \phi}{T_a}. 
	$$
Here, we have denoted the orbit average by an overbar,
	$$ \overline {\left( \cdots \right)} = \int \left( \cdots \right) \frac{dl}{v_\|} 
	   \bigg\slash \int \frac{dl}{v_\|}, 
	$$
where the integrals are taken between consecutive bounce points for trapped orbits and once around the field line for circulating orbits if the field lines close on themselves. If they instead cover a magnetic surface, the integrals are taken over many turns around that surface for circulating orbits. Thus we find that the solution to the gyrokinetic equation becomes, to lowest order,  
	\bn 
	  g_a = \frac{\omega - \omega_{\ast a}^T}{\omega - \overline \omega_{da}} \frac{e_a f_{a0}}{T_a} 
	  \left( \overline{\phi } - \overline{\psi} + \frac{\overline{\omega_{da} \psi}}{\omega} \right) 
	  + \left( 1 - \frac{\omega_{\ast a}^T}{\omega} \right) \frac{e_a \psi}{T_a} f_{a0}
	\label{g0}
	\en

Our next step is to derive a system of two equations for $\psi$ and $\phi$. As in %Ref.~
\cite{Tang-1980} we operate with
	$$ \sum_a e_a \int \left( \cdots \right) J_0 d^3v 
	$$
on the gyrokinetic equation (\ref{gk}). The first term then becomes
	$$ \frac{B}{\mu_0 \omega} \nabla_\| \left( \frac{k_\perp^2}{B} \nabla_\| \psi \right)
	   - i \sum_a e_a \int v_\| g_a \nabla_\| J_0, 
	$$
where we have used Eq.~(\ref{A}) and shall neglect $\nabla_\| J_0$ on the grounds that $b \ll 1$, and the second term becomes
	$$ \frac{2 n e^2 \omega}{T} \left(1 + k_\perp^2 \lambda_D^2 \right) \phi
	   - \sum_a e_a \int \omega_{da} g_a d^3v, 
	$$
where we have used Poisson's law (\ref{phi}). Substituting the solution (\ref{low-omega}) in the second of these terms and again letting $J_0 \rightarrow 1$ gives
	$$ \sum_a e_a \int \omega_{da} g_a d^3v = \frac{2e^2}{T} \int \omega_d f_0 \left[
	   \frac{\omega (\overline{\omega}_d - \omega_\ast^T )}{\omega^2 - \overline{\omega}_d^2} 
		 \left( \overline{\phi} - \overline{\psi} \right)
	   + \frac{\omega^2 - \overline{\omega}_d \omega_\ast^T}{\omega^2 - \overline{\omega}_d^2} 
		 \frac{\overline{\omega_d \psi}}{\omega}
	   - \frac{\omega_\ast^T \psi}{\omega}\right] d^3v, 
	$$
where quantities without species subscript refer to the positrons, as always. Here
	$$ \int \omega_d \omega_\ast^T f_0 d^3v = n \hat{\omega}_d \omega_p, 
	$$
where we have written $ \omega_p = \omega_\ast (1 + \eta) $ and
	$$ \omega_d = \frac{\hat{\omega}_d}{v_T^2} \left( \frac{v_\perp^2}{2} + v_\|^2 \right), 
	$$
assuming $\beta \ll 1$. Collecting all the terms, we thus obtain the equation
	$$ \frac{B}{\omega^2} \nabla_\| \left( \frac{b v_A^2}{B} \nabla_\| \psi \right)
	   + \frac{\hat{\omega}_d \omega_p}{\omega^2} \psi + k_\perp^2 \lambda_D^2 \phi 
	$$
	\bn
	  = \frac{1}{n} \int \frac{\omega_d f_0}{\omega^2 - \overline{\omega}_d^2}
	  \left[ \left( \overline{\omega}_d - \omega_\ast^T \right) \left(\overline{\phi} - \overline{\psi} \right)
	  + \left(1 - \frac{\overline{\omega}_d \omega_\ast^T}{\omega^2} \right) \overline{\omega_d \psi} 
		\right] d^3v. 
	\label{1}
	\en
A second equation is obtained from Poisson's law (\ref{phi}), which with the distribution functions (\ref{g0}) becomes
	\bn 
	  \left( 1 + k_\perp^2 \lambda_D^2 \right) \phi - \psi = \frac{1}{n}
	  \int \frac{f_0}{\omega^2 - \overline{\omega}_d^2}
	  \left[ \left(\omega^2 - \overline{\omega}_d \omega_\ast^T \right) \left(\overline{\phi} 
		- \overline{\psi} \right)
	  + \left(\overline{\omega}_d - \omega_\ast^T \right) \overline{\omega_d \psi} \right] d^3v. 
	\label{2}
	\en
All low-frequency instabilities in Maxwellian electron-positron plasmas with small gyroradius and beta are governed by these two equations. Because of the symmetry between the two species, these equations only contain the square of $\omega$ and there is no preferred direction of rotation of the eigenmodes. 

\subsection{Low-beta limit}

For the electron-positron plasma experiments mentioned in the Introduction, the pressure is exceedingly small, $\beta \sim 10^{-10}$. The first term in Eq.~(\ref{1}) exceeds the second term by a factor $1/\beta$ if all gradients have a common scale length $L$, and the ratio between the first and third terms is of order
	$$ \frac{B}{\omega^2} \nabla_\| \left( \frac{b v_A^2}{B} \nabla_\| \psi \right)
	   \bigg\slash (k_\perp^2 \lambda_D^2 \phi) \sim \left( \frac{c}{\omega L} \right)^2 \frac{\psi}{\phi}. 
	$$
	
There are now two possibilities: either $\psi \ll \phi$ or $\psi$ is at least as large as $\phi$. In the former case, $\psi$ can be neglected in Poisson's law~(\ref{2}), which reduces to its counterpart in the electrostatic limit. Electromagnetic effects are then unimportant, any instabilities are electrostatic, and the analysis of %Ref.~
\cite{Helander-2014} is sufficient. 

In the case that $\psi$ is not much smaller than $\phi$, it is clear that the first term in Eq.~(\ref{1}) exceeds all the others. Expanding 
	$$\psi = \psi_0 + \psi_1 + \cdots, 
	$$
	$$ \phi = \phi_0 + \phi_1 + \cdots, 
	$$
accordingly [i.e., in the small parameter $(\omega / cL)^2$], we must require $\nabla_\| \psi_0 = 0$, i.e., we only consider flute/interchange modes. These are only possible if the magnetic field lines close upon themselves. (If they do not, there are no electromagnetic instabilities at the low beta values we consider. Ballooning modes are thus absent.) Furthermore, since $\psi_0 = \overline \psi_0$ 
Poisson's law~(\ref{2}) reduces to its electrostatic counterpart,
	$$ \left( 1 + k_\perp^2 \lambda_D^2 \right) \phi_0 = \frac{1}{n}
	   \int \frac{\omega^2 - \overline{\omega}_d \omega_\ast^T}{\omega^2 - \overline{\omega}_d^2}
	   \overline{\phi}_0 f_0 d^3v, 
	$$
There are two ways of satisfying this equation, depending on the magnitude of the frequency $\omega$. If it is comparable to $\omega_d$ and $\omega_\ast$, we need to find a non-trivial solution, i.e., an electrostatic mode is established and the problem again reduces to that considered by
\cite{Helander-2014}. Alternatively, if $\omega$ is much larger than the diamagnetic and drift frequencies, and in addition $k_\perp \lambda_D \ll 1$, there is a ``trivial'' solution, namely, 
$\phi_0 = $ constant. (The condition $k_\perp \lambda_D \ll 1$ is necessary since $k_\perp$ varies along the magnetic field.) The frequency $\omega$ then remains undetermined in this order, and we need to proceed to the next order of the expansion of Eq.~(\ref{1}), which becomes
	$$ \frac{B}{\omega^2} \nabla_\| \left( \frac{b v_A^2}{B} \nabla_\| \psi_1 \right) 
	   + k_\perp^2 \lambda_D^2 \phi_0 = \frac{1}{n}
	   \int \frac{\omega_d \left( \overline{\omega}_d - \omega_\ast^T \right)}{\omega^2 
		 - \overline{\omega}_d^2} {\phi}_0 f_0 d^3v, 
	$$
where the right-hand side may be approximated by 
	$$ \frac{\phi_0}{n\omega^2}
	   \int \omega_d \left( \overline{\omega}_d - \omega_\ast^T \right) f_0 d^3v. 
	$$
The integrability condition for this equation, which results from multiplication by $dl/B$ and integration around the closed field line, determines the frequency,
	\bn 
	  \omega^2 = \frac{1}{n} \oint \frac{dl}{B}
	  \int \omega_d \left( \overline{\omega}_d - \omega_\ast^T \right) f_0 d^3v
	  \bigg\slash \oint \left(k_\perp \lambda_D \right)^2 \frac{dl}{B}, 
	\label{interchange growth rate}
	\en
which is indeed smaller than $v_T/L$ but larger than $\omega_d$ and $\omega_\ast$, as required by the orderings. This instability is the usual electromagnetic interchange mode, modified by Debye shielding.

\subsection{Conventional ballooning modes}

Although ballooning modes, familiar from tokamaks and stellarators, cannot appear at the extremely low pressures we are primarily interested in, it is perhaps of interest to see how these modes arise in the present formalism. Ballooning modes are characterised by long, but finite, wavelength along the magnetic field, and are found when the perpendicular wavelength exceeds the Debye length, $k_\perp \lambda_D \ll 1$, and the frequency exceeds the diamagnetic frequency, $\omega_d \sim \omega_\ast \ll \omega$, so that 
Eq.~(\ref{2}) reduces to 
	$$ \phi - \psi = \frac{1}{n} \int f_0 \left(\overline{\phi} - \overline{\psi} \right) d^3v. 
	$$
Since the right-hand side does not depend on the coordinate along the magnetic field, the difference $\phi - \psi$ is constant along the field, and the parallel electric field vanishes. Thus, $\phi = \psi$ within an unimportant integration constant, and Eq.~(\ref{1}) reduces to 
	\bn 
	  B \nabla_\| \left( \frac{b v_A^2}{B} \nabla_\| \psi \right)
	  + \left( \hat{\omega}_d \omega_p  + \omega^2 k_\perp^2 \lambda_D^2 \right) \psi 
	  = \frac{1}{n} \int \omega_d \overline{\omega_d \psi} f_0 d^3v. 
	\label{interchange modes}
	\en
A quadratic form for the eigenvalue $\omega^2$ is obtained by multiplying by $\psi dl/B$ and integrating along the field line,
	\bn 
	  \omega^2 = \frac{N[\phi]}{D[\phi]}, 
	\label{variational form}
	\en
with
	$$ N[\phi] = \int \left[ b v_A^2\left( \frac{\p \psi}{\p l} \right)^2 
	   - \hat{\omega}_d \omega_p \psi^2 \right] \frac{dl}{B}
	   + \frac{1}{n} \int  f_0 \left( \overline{\omega_d \psi} \right)^2 d^3v, 
	$$
	$$ D[\phi] = \int \left( k_\perp \lambda_D \psi \right)^2 \frac{dl}{B}. 
	$$
The form~(\ref{variational form}) is variational, i.e., if a test function $\phi$ is used, then the right-hand side assumes its lowest value for the actual function $\phi$ solving the eigenvalue problem 
(\ref{interchange modes}). 

A few properties of the eigenmodes are immediately clear from the variational form. First, $\omega^2$ is always real, so that the eigenmodes are either oscillatory or purely growing/damped, just like in ideal MHD. Second, it is clear that an instability can only arise if the product $\omega_d \omega_p$ is positive somewhere along the magnetic field. This is, of course, the condition of unfavorable magnetic curvature. Otherwise, $N[\phi]$ is positive and $\omega^2 > 0$. Third, in a closed-line system, a possible test function is $\psi = 1$, and the variational principle implies that $\omega^2$ is less than, or equal to, the interchange-mode result~(\ref{interchange growth rate}). Note that the estimate $\omega \sim \omega_p / (k_\perp \lambda_D)$ implies that $\omega$ indeed exceeds $\omega_d$ but is smaller than the bounce frequency, as assumed in our orderings. 

\section{The case of a dipole field}
	
\begin{figure}
  \centerline{\includegraphics[width=0.7\textwidth]{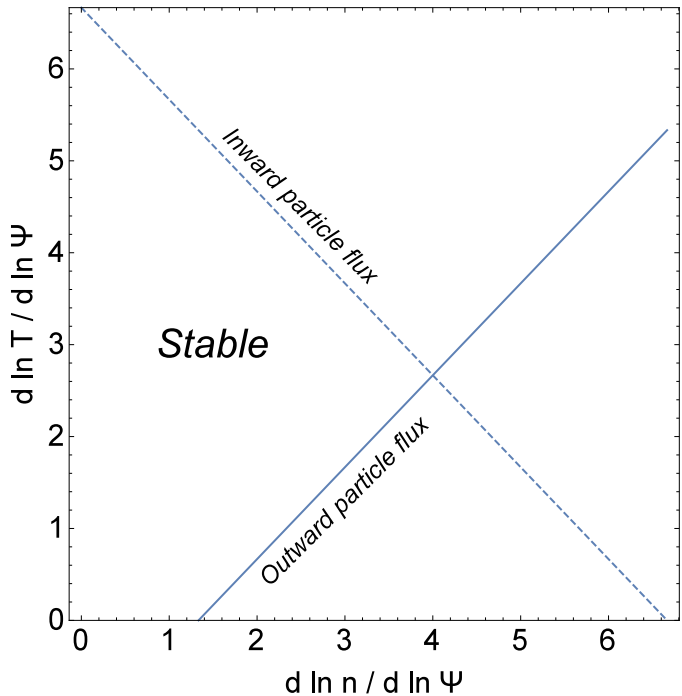}}% Images in 100% size
  \caption{Stability diagram of an electron-positron plasma in a dipole magnetic field. 
	         Electrostatic   modes are unstable below the solid line, and electromagnetic ones 
					 above the dashed line. In the vicinity of the stability boundary, the quasilinear 
					 particle flux is inward if $d \ln T / d \ln n > 2/3$.}
\label{fig1}
\end{figure}

We now specialise to the simplest magnetic geometry for a pair-plasma experiment, that of a magnetic dipole. Such a field may be created by a levitated superconducting coil, as in the Levitated Dipole Experiment (LDX) at MIT (\cite{LDX-2010}), and is of obvious interest for the study of magnetospheric plasmas. For simplicity, we take the limit of a point dipole, where the magnetic field is equal to $ {\bf B} = \nabla \Psi \times \nabla \varphi$, with $\Psi = (M/r) \sin^2 \theta$ in spherical coordinates $(r,\theta,\varphi)$. Note that we have taken $\Psi$ to increase inwards. The volume enclosed by a flux surface is equal to
	$$ V(\Psi) = \frac{16}{35} \frac{4 \pi R^3}{3}, 
	$$
where $R = M/\Psi$ denotes the radius in the equatorial plane. Particles confined in a dipole field precess azimuthally with a frequency that is almost independent of the pitch angle (\cite{Kesner-2002}). The general expression for the latter, the average precession rate, is equal to (\cite{Helander-2014-ROP})
$$ \lang \int_0^1 {\bf v}_d \cdot \nabla \varphi \; d\xi \rang 
	 = - \frac{v^2 B}{3 \Omega } \frac{V''}{V'}, 
$$
where angular brackets denote the flux-surface average (defined as a volume average between neighbouring flux surfaces), and we thus approximate the drift frequency by
		\bn \omega_d = \frac{4 k_\varphi m v^2}{3e\Psi}. 
		\label{precession}
		\en
The stability threshold then obtained from (\ref{interchange growth rate}) becomes
	$$ \frac{d \ln p}{d \ln \Psi} = \frac{20}{3}, $$
and coincides with the stability criterion for interchange modes in ideal MHD that was used in 
%Ref.~
\cite{Helander-2014}. As in that reference, the stability diagram thus looks like that in 
Fig.~\ref{fig1}. 

%\begin{figure}[htb]
%\begin{center}
%\includegraphics[width=0.7\textwidth]{Fig1-JPP.eps}
%\caption{\em Stability diagram of an electron-positron plasma in a dipole magnetic field. Electrostatic modes are unstable below the solid line, and electromagnetic ones above the dashed line. In the vicinity of the stability boundary, the quasilinear particle flux is inward if $d \ln T / d \ln n > 2/3$.}
%\end{center}
%\label{fig1}
%\end{figure}

\section{Quasilinear transport}

A remarkable property of plasma confinement in a dipole magnetic field is that the pressure profile has a natural tendency to become peaked. Turbulent field fluctuations can carry particles inward, up the density gradient. This behaviour was predicted theoretically in connection with planetary magnetospheres 
(\cite{Birmingham-1969}) and has been confirmed experimentally in the laboratory (\cite{LDX-2010,Saitoh-2011-RT-1}), where it provides a robust mechanism for density peaking in levitated dipole fields. Hasegawa, Mauel and Chen proposed a levitated dipole as a configuration suitable for achieving fusion power through the D-$^3$He reaction, and the spontaneous peaking of the pressure profile plays an important role in their proposal (\cite{Hasegawa-1990}). 

A pair-plasma experiment of the nature discussed in the present paper could benefit from inward particle transport, if this made it possible to fuel the plasma from the edge rather than having to inject all particles into the core of the confinement volume. (Magnetic plasma confinement otherwise works equally well both ways: the field is just as good at keeping particles in as out!) It is therefore of interest to calculate the quasilinear particle flux
	$$ \Gamma_a = {\rm Re} \lang \int\frac{f_{a1}}{B} \left({\bf E}^\ast \times {\bf b} 
	   +  v_\| \delta {\bf B}^\ast \right) \cdot \nabla \Psi
	   \; d^3v \rang  = - k_\varphi {\rm Im} \lang \int g_a 
		 \left( \phi^\ast - v_\| A_\|^\ast \right) \; d^3v \rang, 
	$$
where an asterisk denotes the complex conjugate. Since the second term within the brackets is relatively large, it is again helpful to write $A_\| = - i \nabla_\| \psi / \omega$ (which rules out magnetic reconnection, overlapping magnetic islands and Rechester-Rosenbluth diffusion) and integrate by parts, to obtain an expression, 
	$$ \Gamma_a = - k_\varphi {\rm Im} \lang \int \left( \phi^\ast g_a 
	   + \frac{i v_\|}{\omega^\ast} \psi^\ast \nabla_\| g_a \right) \; d^3v \rang, 
	$$
where we can use the gyrokinetic equation (\ref{gk}) to express $\nabla_\| g_a$ in terms of $g_a$, $\phi$ and $\psi$, giving
	$$ \Gamma_a = - k_\varphi {\rm Im} \lang \int \left[ \phi^\ast g_a
	   + \frac{\psi^\ast}{\omega^\ast} \left( \left(\omega - \omega_{\ast a}^T \right) 
		 \frac{e_a \phi f_{a0}}{T} 
	   - \left( \omega - \omega_{da} \right) g_a \right) \right] \; d^3v \rang. 
	$$	
	As we have seen, at the low values of plasma pressure expected in pair-plasma experiments, $\psi$ either vanishes or is almost constant along magnetic field lines. In both cases, Eq.~(\ref{g0}) reduces to its electrostatic counterpart,
	$$ g_a = \frac{\omega - \omega_{\ast a}^T}{\omega - \overline{\omega}_{da}} \; 
	   \frac{e_a \overline{\phi} f_{a0} }{T} , 
	$$
and the quasilinear particle flux becomes
	\bn 
	  \Gamma_a = - \frac{e_a k_\varphi}{T} {\rm Im} \lang \int f_{a0}  
		\frac{\omega - \omega_{\ast a}^T}{\omega - \overline{\omega}_{da}} \;
	  \left| \overline{\phi} \right|^2 \; d^3v \rang. 
	\label{ql_flux}
	\en

To proceed, we would have to solve the eigenvalue problem to determine the frequency $\omega$ and mode structure $\phi(l)$ along the magnetic field, and evaluate the integral in Eq.~(\ref{ql_flux}). We cannot achieve this analytically except close to marginal stability, where $\omega = \omega_r + i \gamma$ with 
$\gamma \rightarrow 0+$ and the resonant denominator results in a delta function, 
	$$ \Gamma_a = \frac{\pi e_a k_\varphi}{T} \lang 
	   \int \delta (\omega - \overline{\omega}_{da}) (\omega - \omega_{\ast a}^T) 
	   \left| \overline{\phi} \right|^2 f_{a0} \; d^3v \rang. 
	$$
The remaining integral is easily evaluated in the approximation (\ref{precession}) that 
$\overline{\omega}_{da}$ is independent of the pitch angle, giving
	$$ \Gamma_a = -\frac{3 \pi^{1/2} n |k_\varphi e_a |}{4 T} \left[ \frac{d \ln n}{d \ln \Psi} 
	   \left( 1 - \frac{3 \eta}{2} + \eta y^2 \right)
	   - \frac{8 y^2}{3} \right] \lang |\phi|^2 \rang y e^{-y^2}, 
	$$
where the resonance condition $\omega = \overline{\omega}_{da}$ implies 
	$$ y^2 = \frac{3 e_a \Psi \omega}{8 k_\varphi T} 
	$$
for dipole geometry. Finally, we recall that $ \omega $ vanishes at marginal stability (\cite{Helander-2014}), so that the flux reduces to
	$$ \Gamma_a = - \frac{3\pi^{1/2} n |k_\varphi e_a y| }{4 T} \frac{d \ln n}{d \ln \Psi} \left( 1 
	   - \frac{3 \eta}{2} \right)
	   \lang |\phi|^2 \rang 
	$$
in its vicinity. This result implies that the particle flux will be in the direction of the density gradient if 
	$$ \eta > \frac{2}{3}. 
	$$
Thus, in the stability diagram of Fig.~1, the particle flux is inward (assuming a centrally peaked density profile, $dn/d\Psi>0$) along much of the stability boundary. In practice, this would mean that an inward particle flux will usually arise if enough heating power is applied to the plasma. (For the extremely tenuous, cold plasmas mentioned in the Introduction, this power would be very small. Of course, one would want to heat the plasma as little as possible to keep the Debye length short.)

\section{Conclusions}

As pointed out in a previous publication
(\cite{Helander-2014}), electron-positron plasmas enjoy remarkable stability properties. These have been explored further in the present paper by including magnetic-field fluctuations in the gyrokinetic equation. As in the electrostatic case, no instabilities arise if the equilibrium magnetic field is constant. Furthermore, if the density is sufficiently low that the Debye length exceeds the gyroradius, all instabilities with wavelengths comparable to the latter are stabilised by Debye screening. In other words, instabilities with $k_\perp \rho = O(1)$, which are thought to drive most of the turbulent transport in fusion devices, are absent since the plasma is too tenuous to support collective motion on such short length scales. Any remaining gyrokinetic instabilities must have frequencies smaller than the bounce (or transit) frequency of the particles. 

Moreover, electromagnetic instabilities are predicted to be absent at the extremely low betas expected in pair-plasma experiments unless the field lines close on themselves. Interchange modes are then destabilised if the logarithmic pressure gradient exceeds a certain threshold, which is possible at arbitrarily small beta. This is analogous to the usual criterion in ideal MHD, which states that $p U^{5/3}$ must not increase in the direction of the curvature vector of the magnetic field lines if the plasma is to be stable against interchange modes (\cite{Rosenbluth-Longmire-1957}). (Here $U$ denotes the specific volume of the magnetic flux tubes.) In magnetic configurations where the precession frequency is independent of pitch-angle, which is approximately the case in the field of a dipole, this criterion coincides with that derived from kinetic theory. The stability diagram from %Ref.~
\cite{Helander-2014}, which is reproduced in Fig.~1, thus remains valid beyond the ideal-MHD approximation. 

Finally, we have considered the quasilinear particle flux and found that it is inward in a dipole plasma close to the marginal stability curve if $\eta > 2/3$, i.e., if the temperature profile is more than 2/3 times steeper than the density profile. This may facilitate the fuelling of a dipole-plasma experiment by causing spontaneous density peaking without the need for a particle source in the plasma core.

\section*{Acknowledgment}

The first author is grateful to Merton College, Oxford, for its splendid hospitality during a visit that lead to the completion of this work.

\bibliographystyle{jpp}

\bibliography{per}

\end{document}